# Open Quantum to classical phases transition in the stochastic hydrodynamic analogy: the explanation of the Lindemann relation and the analogies between the maximum of density at He lambda point and that one at water-ice phase transition

**Piero Chiarelli***

*National Council of Research of Italy, Area of Pisa, 56124 Pisa, Moruzzi 1, Italy*

*and Interdepartmental Center "E.Piaggio" University of Pisa*

***Authors' contributions:***

*This work was carried out by the author that designed the study, performed the calculations and managed the analyses and the literature searches of the study. The author read and approved the final manuscript*

***Received***
***Accepted***
***Published***

ABSTRACT

In the present paper the gas, liquid and solid phases made of structureless particles, are visited to the light of the quantum stochastic hydrodynamic analogy (SQHA). The SQHA shows that the open quantum mechanical behavior is maintained on a distance shorter than the theory-defined quantum correlation length ($\lambda_c$). When, the physical length of the problem is larger than $\lambda_c$, the model shows that the quantum (potential) interactions may have a finite range of interaction maintaining the non-local behavior on a finite distance "quantum non-locality length" $\lambda_q$. The present work shows that when the mean molecular distance is larger than the quantum non-locality length we have a "classical" phases (gas and van der Waals liquids) while when the mean molecular distance becomes smaller than $\lambda_q$ or than $\lambda_c$ we have phases such as the solid crystal or the superfluid one, respectively, that show quantum characteristics. The model agrees with Lindemann empirical law (and explains it), for the mean square deviation of atom from the equilibrium position at melting point of crystal, and shows a connection between the maximum density at the He lambda point and that one near the water-ice solidification point.

## 1. INTRODUCTION

The breaking of quantum coherence is a problem that has many consequences in all problems of physics whose scale is larger than that one of small atoms that are dynamically submitted to environmental fluctuations such as chromophore-protein complexes, semi-conducting polymers and quantum to classical phase transition.
The suitability of the classical-like theories in explaining those open quantum phenomena is confirmed by their success in the description of dispersive effects in semiconductors, multiple tunneling, mesocopic and quantum Brownian oscillators, critical phenomena, stochastic Bose-Einstain condensation [1-11]. The interest for the quantum hydrodynamic analogy (QHA) [12] has been recently growing by its strict relation with the Schrödinger mechanics [13] (resulting useful in the numerical solution of the time-dependent Schrödinger equation [14]) and for the absence of logical problem such the undefined variables of the Bohmian mechanics [15] leading to a number of papers and textbooks bringing original contributions to the comprehension of quantum dynamics [16-19].
Recently the author has developed the stochastic version of the QHA (SQHA) [20]. Such a theory shows that fluctuations of the wave function modulus WFM cannot have a white spatial spectrum (zero correlation distance). Those short-distance wrinkles of the WFM are energetically suppressed in order to maintain the energy of the fluctuating state finite. This quantum noise suppression is the mechanism by which the standard quantum mechanics (the deterministic limit of the SQHA) is realized on short scale dynamics. By imposing the condition of a finite energy for any fluctuating state, the model derives the characteristic distance $\lambda_c$ (named quantum correlation length) below which the standard quantum (non-local) mechanics is achieved.
Moreover, the SQHA analysis also shows that when the inter-particle interaction is weaker than the linear one, it is possible to have a finite range of interaction of the so called quantum pseudo-potential [12,14] (named *quantum non-locality length* $\lambda_q$) with $\lambda_q \geq \lambda_c$, so that on large scale the classical mechanics can be realized.
In this paper we investigate how the existence of a finite range of quantum interaction affects the behavior of a system of a huge number of structureless particles, realizing different physical phases depending by the ratio between $\lambda_q$ and the mean inter-particle distance.
The particle confinement, is also discussed to the light of the SQHA model and shown how it can be achieved in a gas phase.

## 2. THEORY: THE SQHA EQUATION OF MOTION

When the noise is a stochastic function of the space, in the SQHA the motion equation is described by the stochastic partial differential equation (SPDE) for the spatial density of number of particles n (i.e., the wave function modulus squared (WFMS)), that reads [20]

$$\partial_t n_{(q,t)} = -\nabla_q \bullet ( n_{(q,t)} \dot{q} ) + \eta_{(q,t,\Theta)} \quad (1)$$

$$< \eta_{(q_\alpha)}, \eta_{(q_\beta + \lambda)} > = < \eta_{(q_\alpha)}, \eta_{(q_\alpha)} > G(\lambda) \delta_{\alpha\beta} \quad (2)$$

$$\dot{p} = -\nabla_q (V_{(q)} + V_{qu(\mathrm{n})}),\tag{3}$$

$$\dot{q} = \frac{\nabla_q S}{m} = \frac{p}{m},\tag{4}$$

$$S = \int_{t_0}^{t} dt \left( \frac{p \bullet p}{2m} - V_{(q)} - V_{qu(\mathrm{n})} \right)\tag{5}$$

where $\Theta$ is the amplitude of the spatially distributed noise $\eta$, $V_{(q)}$ represents the Hamiltonian potential and $V_{qu(\mathrm{n})}$ is the so-called (non-local) quantum potential [12,14] that reads

$$V_{qu} = -\left(\frac{\hbar^2}{2m}\right)\mathrm{n}^{-1/2}\nabla_q \bullet \nabla_q \mathrm{n}^{1/2}.\tag{6}$$

Moreover, $G(\lambda)$ is the dimensionless shape of the correlation function of the noise $\eta$.
The condition that the fluctuations of the quantum potential $V_{qu(\mathrm{n})}$ do not diverge, as $\Theta$ goes to zero (so that the deterministic limit can be warranted) leads to a $G(\lambda)$ owing the form [20]

$$\lim_{\Theta \to 0} G(\lambda) = exp\left[-\left(\frac{\lambda}{\lambda_c}\right)^2\right].\tag{7}$$

This result is a direct consequence of the quantum potential form that owns a membrane elastic-like energy, where higher curvature of the WFMS leads to higher energy. White fluctuations of the WFMS that bring to a zero curvature wrinkles of the WFMS (and hence to an infinite quantum potential energy) are not allowed. In order to maintain the system energy finite independent fluctuations are progressively suppressed on shorter and shorter distance, leading, in the small noise limit, to the existence of a correlation distance (let's name it $\lambda_c$) for the noise.
Thence, (2) reads [20]

$$\lim_{\Theta \to 0} < \eta_{(q_\alpha, t)}, \eta_{(q_\beta + \lambda, t+\tau)} > = \underline{\mu}\frac{8m(k\Theta)^2}{\pi^3 \hbar^2} exp\left[-\left(\frac{\lambda}{\lambda_c}\right)^2\right]\delta(\tau)\delta_{\alpha\beta}\tag{8}$$

where

$$\lim_{\Theta \to 0} \lambda_c = \left(\frac{\pi}{2}\right)^{3/2} \frac{\hbar}{(2mk\Theta)^{1/2}}\tag{9}$$

and where $\underline{\mu}$ is the WFMS mobility form factor that depends by the specificity of the considered system [20].

### 2.1 Range of interaction $\lambda_q$ of quantum pseudo-potential

In addition to the noise correlation function (7), in the large-scale limit, it is also important to know the behavior of the quantum force $\dot{p}_{qu} = -\nabla_q V_{qu}$ at large distance.

The relevance of the force generated by the quantum potential at large distance can be evaluated by the convergence of the integral [20]

$$\int_0^{\infty} | q^{-1} \nabla_q V_{qu} | dq \qquad (10)$$

If the quantum potential force at large distance grows less than a constant (so that $\lim_{|q|\to\infty} | q^{-1} \nabla_q V_{qu} | \propto | q |^{-(1+\varepsilon)}$, where $\varepsilon$ > 0) the integral (10) converges. In this case, the mean weighted distance

$$\lambda_q = 2 \frac{\int_0^{\infty} | q^{-1} \frac{dV_{qu}}{dq} | dq}{\lambda_c^{-1} | \frac{dV_{qu}}{dq} |_{(q=\lambda_c)}}, \qquad (11)$$

can evaluate the quantum potential range of interaction.

Faster the Hamiltonian potential grows, more localized is the WFMS and hence stronger is the quantum potential. For the linear interaction, the Gaussian-type eigenstates leads to a quadratic quantum potential (see section 3.2) and, hence, to a linear quantum force, so that $\lim_{|q|\to\infty} | q^{-1} \nabla_q V_{qu} | \propto constant$ and $\lambda_q$ diverges. Therefore, in order to have $\lambda_q$ finite (so that the large-scale classical limit can be achieved) we have to deal with a system of particles interacting by a force weaker than the linear one.

## 2.2 Scale-depending SQHA dynamics

1) *Non-local deterministic dynamics* (i.e., the standard quantum mechanics)

$\Delta L << \lambda_c \cup \lambda_q$ (e.g., $\Theta \to 0$) where $\Delta L$ is the characteristic physical length of the problem leads to

$$\partial_t \mathrm{n}_{(q,t)} = -\nabla_q \bullet ( \mathrm{n}_{(q,t)} \dot{q} ) \qquad (12)$$

That is equivalent to the Schrödinger equation [25].

2) *Non-local stochastic dynamics*, with $\lambda_c < \Delta L << \lambda_q$

$$\partial_t \mathrm{n}_{(q,t)} = -\nabla_q \bullet ( \mathrm{n}_{(q,t)} \dot{q} ) + \eta_{(q,t,\Theta)} \qquad (13)$$

$$< \eta_{(q_\alpha,t)}, \eta_{(q_\beta+\lambda,t+\tau)} > = \underline{\mu} \, \delta_{\alpha\beta} \frac{2k\Theta}{\lambda_c} \delta(\lambda) \delta(\tau) \qquad (14)$$

3) *Local stochastic dynamics*, with $\lambda_c \le \lambda_q << \Delta L$.

Given the condition $\lambda_q << \Delta L$ so that it holds

$$\lim_{q\to\infty} -\nabla_q V_{qu(\mathrm{n}_0)} = 0 \qquad (15)$$

the SPDE of motion acquires the form

$$\partial_t n_{(q,t)} = -\nabla_q \bullet (n_{(q,t)} \dot{q}) + \eta_{(q,t,\Theta)} \tag{16}$$

$$< \eta_{(q_\alpha ,t)}, \eta_{(q_\beta + \lambda, t+\tau)} > = \underline{\mu}\, \delta_{\alpha\beta} \frac{2k\Theta}{\lambda_c} \delta(\lambda)\delta(\tau) \tag{17}$$

$$\begin{aligned} \dot{q} &= \frac{p}{m} = \nabla_q \lim_{\Delta L / \lambda_q \to \infty} \frac{\nabla_q S}{m} = \nabla_q \{ \lim_{\Delta L / \lambda_q \to 0} \frac{1}{m} \int_{t_0}^{t} dt( \frac{p \bullet p}{2m} - V_{(q)} - V_{qu} ) \} \\ &= \frac{1}{m} \{ \int_{t_0}^{t} dt\, ( \frac{p \bullet p}{2m} - \nabla_q V_{(q)} - \Delta ) \} = \frac{p_{cl}}{m} + \frac{\delta p}{m} \cong \frac{p_{cl}}{m} \end{aligned} \tag{18}$$

where

$$\Delta = \lim_{\Delta L / \lambda_q \to 0} \nabla_q ( V_{qu(n)} - V_{qu(n_0)} ), \tag{19}$$

$\delta p$ is a small fluctuation of momentum and

$$\dot{p}_{cl} = -\nabla_q V_{(q)}. \tag{20}$$

## 3. QUANTUM BEHAVIOR OF PSEUDO-GAUSSIAN FREE PARTICLES IN THE DETERMINISTIC SQHA LIMIT

In order to elucidate the interplay between the Hamiltonian potential and the quantum potential, that together define the evolution of the particle wave function modulus (WFM), we observe that the quantum potential is primarily defined by the WFM.

Fixed the WFM at the initial time, then the Hamiltonian potential and the quantum one determine the evolution of the WFM that on its turn modifies the quantum potential.

A Gaussian WFM has a parabolic repulsive quantum potential, if the Hamiltonian potential is parabolic too (the free case is included), when the WFM wideness adjusts itself to produce a quantum potential that exactly compensates the force of the Hamiltonian one, the Gaussian states becomes stationary (eigenstates). In the free case, the stationary state is the flat Gaussian (with an infinite variance) so that any Gaussian WFM expands itself following the ballistic dynamics of quantum mechanics since the Hamiltonian potential is null and the quantum one is quadratic [see Appendix A].

From the general point of view, we can say that if the Hamiltonian potential grows faster than a harmonic one, the wave equation of a self-state is more localized than a Gaussian one and this leads to a stronger-than a quadratic quantum potential (at large distance).

On the contrary, a Hamiltonian potential that grows slower than a harmonic one will produce a less localized WFM that decreases slower than the Gaussian one [see Appendix A ], so that the quantum potential is weaker than the quadratic one and it may lead to a finite quantum non-locality length (11).

More precisely, the large distances exponential-decay of the WFM such as

$$m\text{-}s \lim_{|q| \to \infty} n^{1/2} \propto exp[-P^{h}{}_{(q)}] \tag{21}$$

with $h < 3/2$

is a sufficient condition to have a finite quantum non-locality length [20].

In absence of noise, we can enucleate three typologies of quantum potential interactions ( in the unidimensional case):

(1) $h > 2$ strong quantum potential that leads to quantum force that grows faster than linearly and $\lambda_q$ is infinite (*super-ballistic* free particle WFM expansion) and reads

$$\lim_{|q|\to\infty} \frac{dV_{qu}}{dq} \propto q^{1+\varepsilon}. \qquad (\varepsilon>0) \qquad (22)$$

(2) $h = 2$ that leads to quantum force that grows linearly (i.e., $\frac{dV_{qu}}{dq} \propto q$ ) and $\lambda_q$ is infinite (*ballistic* free WFM enlargement)

$$\lim_{|q|\to\infty} \frac{dV_{qu}}{dq} \propto q \qquad (23)$$

(3) $2 > h \geq 3/2$ "middle quantum potential";
the integrand of (11) will result

$$Const > \lim_{|q|\to\infty} | q^{-1} \frac{dV_{qu}}{dq} | > q^{-1}. \qquad (24)$$

The quantum force grows less that linearly at large distance but $\lambda_q$ may be still infinite (*under-ballistic* free WFM expansion).
(4) $h < 3/2$ "week quantum potential" interaction leading to quantum force that becomes vanishing at large distance following the asymptotic behavior

$$\lim_{|q|\to\infty} | q^{-1} \frac{dV_{qu}}{dq} | > q^{-(1+\varepsilon)}, (\varepsilon > 0) \qquad (25)$$

with a finite $\lambda_q$ for $\Theta \neq 0$ (*asymptotically vanishing* free WFM expansion).

### 3.1. Free pseudo-Gaussian particles in presence of noise

Gaussian particles generate a quantum potential that has an infinite range of interaction and hence cannot lead to macroscopic local dynamics.
Nevertheless, imperceptible deviation by the perfect Gaussian WFM may possibly lead to finite quantum non-locality length [see Appendix A]. Particles that are inappreciably less localized than the Gaussian ones (let's name them as pseudo-Gaussian) own $\frac{dV_{qu}}{dq}$ that can sensibly deviate by the linearity so that the quantum non-locality length may be finite.
In the case of a free pseudo-Gaussian particle we can say that $\lambda_q$ extends itself at least up to the Gaussian core $(\underline{\Delta q^2})^{1/2}$ (where the quantum force is linear). At a distance much bigger than $(\underline{\Delta q^2})^{1/2}$ for $h < 3/2$, the expansive quantum force becomes vanishing.

Taking also into account that on short distance, for q << $\lambda_c$, the noise is progressively suppresses (i.e., the deterministic quantum dynamics is established), it follows that:

(1) For $(\underline{\Delta q}^2)^{1/2} << \lambda_c$, the expansion dynamics of the free pseudo-Gaussian WFM are almost ballistic (quantum deterministic) .

(2) For $(\underline{\Delta q}^2)^{1/2} >> \lambda_q$ and for *h* < 3/2 the expansion dynamics of the free pseudo-Gaussian WFM are almost diffusive.

For $\lambda_c << (\underline{\Delta q}^2)^{1/2} < \lambda_q$ the noise will add diffusion to the WFM ballistic enlargement.

When the (pseudo-Gaussian) WFM has reached the mesoscopic scale ($(\underline{\Delta q}^2)^{1/2} \sim \lambda_q$), we have that its core expands ballistically while its tail diffusively.

Since the outermost expansion is slower than the innermost, there is an accumulation of WFM (that is a conserved quantity) in the middle region (q ~ $\lambda_q$) generating, as time passes, a slower and slower (than the Gaussian one) WFM decrease so that (for a free particle) *h*, as well as the quantum potential and $\lambda_q$ decrease (and cannot increase) in time.

In force of these arguments (i.e., the quantum ballistic core enlargement faster than the classical diffusive peripheral one), the free pseudo-Gaussian states (with *h* < 3/2) is a self-sustained state and remain pseudo-Gaussian in time.

As far as it concerns the particle de-localization at very large times, the asymptotically vanishing quantum potential does not completely avoid such a problem since the (Θ-noise diffusion driven) spreading of the molecular WFM remains (even if it is much slower than the quantum ballistic one).

If the particle WFM confinement cannot be achieved in the case of one or few molecules, on the contrary, in the case of a system of a huge number of structureless particles (with a repulsion core as in the case of the Lennard-Jones (L-J) potentials) the WFM localization comes from the interaction (collisions) between the molecules.

More analytically, we can say that in a rarefied gas phase, when two colliding particles get at the distance of order of the L-J potential minimum $r_0$, the quantum non-locality length becomes sensibly different from zero and bigger than the inter-particle distance $r_0$ since (for a sufficiently deep L-J well) the potential is approximately quadratic and the associated state has a Gaussian core (i.e., pseudo-Gaussian)).

After the collision, when the molecules are practically free, the pseudo-Gaussian WFM again starts to freely expand leading to a new decrease of $\lambda_q$.

In this way, $\lambda_q$ will never reach the zero value since, in a finite time, the molecule undergoes another collision and maintains a non –zero mean $\lambda_q$ value. This because at the collision the WFM takes a bit of squeezing that leads to a new increase of the $\lambda_q / (\underline{\Delta q}^2)^{1/2}$ ratio.

In this way the WFM will never reach the (free) flat Gaussian asymptotical configuration.

The overall effect of this process is that the random collisions between free particles in a gas phase with L-J type intermolecular potential, maintain their localization.

### 3.2 Quantum non-locality length of L-J bounded states

In order to calculate the quantum potential and its non-locality length for a L-J potential well, we can assume the harmonic approximation of the L-J potential around the reduced equilibrium position $\underline{q} = ½\, r_0$ (where $r_0$ is the molecular distance) that reads

$$V_{L-J} \cong \frac{k}{2}(q-\underline{q})^2 - \upsilon\,, \qquad (26)$$

Where $\upsilon$ is the L-J well deepness and

$$k = \frac{4(E_0+\upsilon)^2 m}{\hbar^2} = \upsilon\left(\frac{12}{r_0}\right)^2 \qquad (27)$$

where $E_0$ is the energy of fundamental state.

Moreover, the convex quadratic quantum potential associated to the wave function

$$\psi_0 = B\,exp[-K_0{}^2(q-\underline{q})^2] \qquad |q-\underline{q}| < \Delta \qquad (28)$$

where

$$K_0 = \frac{((E_0+\upsilon)m)^{1/2}}{\hbar} \qquad (29)$$

reads

$$\begin{aligned} V_{qu} &= -\left(\frac{\hbar^2}{2m}\right)|\psi|^{-1}\nabla_q \bullet \nabla_q |\psi| = -\left(\frac{\hbar^2}{2m}\right)K_0{}^4(q-\underline{q})^2 + (E_0+\upsilon) \\ &= -\frac{k}{2}(q-\underline{q})^2 + (E_0+\upsilon) \end{aligned} \qquad (30)$$

leading to the quantum force

$$-\nabla_q V_{qu} = k(q-\underline{q}) \qquad (31)$$

and to

$$E_0 = V_{L-J} + V_{qu}\,. \qquad (32)$$

For $q > \underline{q} + \delta$, where

$$\delta = r_0 - r_{(V_{l-J}=0)} = 0{,}11785\, r_0 \qquad (33)$$

we can assume that the L-J is approximately a constant leading to an exponential decrease of WFM [*h*=1 of section 3] and hence, to a vanishing small quantum force that we can disregard in the calculus of the quantum non-locality length .

Thence, by (11) and (31) it follows that

$$\lambda_q \cong 2\lambda_c \frac{\int_{\underline{q}}^{\underline{q}+\delta} | q^{-1} \frac{dV_{qu}}{dq} | dq}{| \frac{dV_{qu}}{dq} |_{(q-\underline{q}=\lambda_c)}} = 2\delta = 0{,}23570\, r_0 \qquad (34)$$

The value of $\lambda_q$ calculated by using wave functions with higher energy eigenvalues $E_n$ leads to similar result since the quantum potential is normalized to the wave function modulus (see (6)) and the wave functions are Gaussian.
In order to have the quantum behavior before reaching the melting point, we must have that the wave function dispersion must be smaller than the value $\lambda_q = 2\delta$ and that equals it at the melting point.
Considering that at melting point the atoms of fluid are in a classical phase (a statistical collection of distinguishable couples of interacting atoms) in order to evaluate the variance of the atomic distance, following the approach in ref. [21], we take under consideration the state of a couple of atoms in the n$^{th}$ level of energy $E_n$ equating the mean energy of the melting temperature such as $E_n =< E >_{(T_m)}$.
The variance $\Delta\psi_n$ of the wave function, with eigenvalue $E_n$, is not exactly the mean wave function variance (on the statistical ensemble of distinguishable couples of interacting atoms) that on its turn is an evaluation of the variance of the atomic distance in solids $< \underline{q}^2 >^{1/2}$ when we perform the measurement (just for the correspondence between the modulus of the quantum wave function and the outputs of position measurements). Nevertheless, in order to give an approximated evaluation of the variance of the atomic distance in solids at melting point we equate the variance $\Delta\psi_n$ to the quantum non-locality length $\lambda_q$ to obtain

$$< \underline{q}^2 >^{1/2} \cong \Delta\psi_n \cong \lambda_q \cong 2\delta = 0{,}23570\, r_0$$

the result (34) well agrees with the Lindemann empirical law that sees $< \underline{q}^2 >^{1/2}$ to range between 0,2 and 0,25 times $r_0$ at melting point [21].
Finally, it must be noted that, on the author knowledge, this appears to be the first explanation of the wide verified Lindemann empirical relation.

### 3.3 Quantum coherence length at the fluid-superfluid transition

For small potential well, the liquid phase can realize itself down to a very low temperature [22]. In this case, even if $\lambda_q$ may result smaller than the inter-particle distance (so that the liquid phase is maintained), decreasing the temperature, and hence the amplitude of fluctuations $\Theta$, when $\lambda_c$ grows and becomes of order of the mean molecular distance, the liquid properties, depending by the molecular interaction such as the viscosity, acquire quantum characteristics.
The fluid-superfluid transition can happen if the temperature can be lowered up to the transition point before the solid phase takes place (i.e., very small L-J potential well such as that one of the 4He).
In the following, we applies the SQHA model to the $He_I$->$He_{II}$ transition by using the diatomic He-He square well potential approximation

$$V_{He-He} = \infty \qquad (q < \sigma) \tag{35}$$

$$V_{He-He} = -0{,}82\ \upsilon \qquad (\sigma < q < \sigma + 2\Delta) \tag{36}$$

$$V_{He-He} = 0 \qquad (q > \sigma + 2\Delta), \tag{37}$$

where [23]

$$r_0 = \sigma + \Delta = 7{,}9\,Bohr \tag{38}$$

$$\Delta = 1{,}54 \times 10^{-10} \cong 2{,}9\ Bohr \tag{39}$$

$$\upsilon = 10{,}9 k_b = 1{,}5 \times 10^{-22} J \tag{40}$$

with the wave function [23]

$$\psi_0 = Bsen[K_0(q-\sigma)] \qquad \text{for } |q - \underline{q}| < \frac{\pi}{2K_0} \tag{41}$$

with a eigenvalue $E_0 = -5{,}19 k_b = -7{,}16 \times 10^{-23} j$ (42)

Thence, the quantum potential and quantum force respectively read

$$V_{qu} = -{K_0}^2 \qquad | q - \underline{q} | < \frac{\pi}{2K_0} \tag{43}$$

$$\nabla_q V_{qu} = 0 \qquad | q - \underline{q} | < \frac{\pi}{2K_0} \tag{44}$$

Due to the discontinuities of the square well He-He potential, formula (34) is not useful. If we consider the harmonic approximation for the He-He potential we end with (34) where

$$\lambda_q \cong 2\delta = 0{,}23570\, r_0 < 2\Delta \cong 0{,}4340 r_0 \tag{45}$$

Moreover, due to the very low deepness of the He-He well, there exists just one bonded state (the ground state) [24] so that, even in the ground state the WFM is widely de-localized in the non linear range of L-J potential. As a consequence of that, the quantum potential range of action $\lambda_q$ can be even smaller than in (45).

Since the quantum potential range of interaction is not meaningful if smaller than $\lambda_c$ [20] because the quantum behavior is established anyway on a length smaller than $\lambda_c$, at the He-He lambda point, hence, we must consider $\lambda_c \cong 2\Delta$ as the condition that maintains the quantum behavior for the He-He system and that in (9) leads to

$$\Theta \cong 2{,}17°K \tag{46}$$

The classical to quantum transition in 4He does not come from the linearity of the inter-atomic force (at a distance of order of $\lambda_q$ as in a solid crystal) but comes by the increases of $\lambda_c$ (due to the decrease of amplitude of fluctuations) that becomes of order of the potential well wideness where the wave function is localized.

This happens since the very small deepness of 4He L-J potential does not lead to a solid quantum 4He crystal before the superfluid transition.

## Discussion and conclusion

The SQHA approach shows that the quantum superposition of states does not survive on large scale in presence of fluctuations and when the inter-particle non-linear interaction is weaker than the linear one such as that one given by the Lennard-Jones inter-atomic potential.

In this paper we have evaluated if this hypothesis leads to realistic consequences when we pass from a rarefied to a condensed phases where the inter-molecular distance becomes smaller than the range of quantum non-local interaction.

Fluids and gas phase do not show quantum characteristics while solids and super-fluids give clear evidences of the existence of quantum mechanics.

Solid crystals as well as super-fluids show properties (depending by the molecular characteristics) that do not agree with classical laws.

Here the transition from the solid crystal phase to the (classical) liquid one has been evaluated by using the SQHA model.

The model agrees with Lindemann empirical law for the mean square deviation of atom from the equilibrium position at melting point of crystal. Moreover, the SQHA furnishes a satisfactory explanation of the Lindemann relation that was unexplained by nowadays theories.
When applied to the fluid-superfluid 4He transition, the model also shows that the transition is due to the restoration of quantum (non-local) potential interaction. At the lambda point transition temperature of 2,17 °K the quantum coherence length $\lambda_c$ becomes of order of the wideness of 4He potential well.
The common quantum origin between the $4He_{I}$->$4He_{II}$ fluid-superfluid transition and the fluid-solid one, suggested by the SQHA model, is quite interesting because is able to furnish an interesting explanation of the analogy between the maximum of density at the $4He_{I}$->$4He_{II}$ fluid-superfluid transition and that one of the water-ice phase transition.
At the $4He_{I}$->$4He_{II}$ fluid-superfluid transition the maximum of density has been shown to be produced by the appearance of the repulsive quantum potential interaction [23].
The maximum is produced by the speed of strengthening of quantum potential (causing expansion) and the speed of thermal shrinking of liquid helium during the cooling process toward the superfluid state. This is not in contradiction of the wider accepted explanation that accounts for the maximum density at lambda point to the quantum kinetic energy. In fact, the so called quantum potential of the Madelung approach (6), written in terms of quantum operator, reads

$$V_{qu} = -(\frac{\hbar^2}{2m}) \mathrm{n}^{-1/2} \nabla_q \bullet \nabla_q \mathrm{n}^{1/2} \rightarrow |\psi|^{-1} \frac{p^2}{2m} |\psi|$$

revealing its “*kinetic*” origin. This is an additional evidence that the quantum hydrodynamic analogy and the Schrödinger one do not contradict each other [13,25].
Therefore, being the solid-fluid transition produced by the appearance of quantum potential interaction, it becomes evident that the maximum density at the water-ice transition is generated by the same mechanics confirming what the finest experimentalists have believed for decades. This hypothesis, suggests that others maxima of density at solid-fluid transition may exist when the thermal shrinking of the material is smaller than the corresponding inter-molecular expansion generated by the quantum pseudo-potential.
The SQHA shows that both the linearity of particle interaction and the reduction of amplitude of stochastic fluctuations elicit the emergence of quantum behavior.
The SQHA model shows that the non linearity of physical forces, other than to play an important role in the establishing of thermodynamic equilibrium, is a necessary condition to pass from the quantum to the classical phases and that fluctuations alone are not sufficient for obtaining that.

## REFERENCES

1. Gardner CL. The quantum hydrodynamic model for semiconductor devices. SIAM J. Appl. Math. 1994; 54, 409.
2. Bertoluzza S, and Pietra P. Space-Frequency Adaptive Approximation for Quantum Hydrodynamic Models. Reports of Institute of Mathematical Analysis del CNR, Pavia, Italy, 1998.
3. Jona Lasinio G, Martinelli F, and Scoppola E. New Approach to the Semiclassical Limit of Quantum Mechanics. Comm. Math. Phys. 1981; 80, 223.
4. Ruggiero P, and Zannetti M. Microscopic derivation of the stochastic process for the quantum Brownian oscillator. Phys. Rev. A. 1983; 28, 987.
5. Ruggiero P, and Zannetti, M. Critical Phenomena at *T*=0 and Stochastic Quantization. Phys. Rev. Lett. 1981; 47, 1231;

6. Ruggiero P, and Zannetti M. Stochastic Description of the Quantum Thermal Mixture. Phys. Rev. Lett.1982; 48(15), 963.
7. Ruggiero P, and Zannetti M. Quantum-classical crossover in critical dynamics. Phys. Rev. B. 1983; 27, 3001.
8. Breit JD, Gupta S, and Zaks A. Stochastic quantization and regularization. Nucl. Phys. B. 1984; 233, 61;
9. Bern Z, Halpern MB, Sadun L, and Taubes C. Continuum regularization of QCD. Phys. Lett. 1985;165 B, 151.
10. Ticozzi F, Pavon M, On Time Reversal and Space-Time Harmonic Processes for Markovian Quantum Channels, arXiv: 0811.0929 [quantum-phys] 2009.
11. Morato LM, Ugolini S. Stochastic Description of a Bose–Einstein Condensate. Annales Henri Poincaré. 2011;12(8), 1601-1612.
12. Madelung E. Quanten theorie in hydrodynamische form (Quantum theory in the hydrodynamic form). Z. Phys. 1926; 40, 322-6, German.
13. Tsekov R. Bohmian mechanics versus Madelung quantum hydrodynamics. arXiv:0904.0723v8 [quantum-phys] 2011;
14. Weiner JH, Askar A. Particle Method for the Numerical Solution of the Time-Dependent Schrödinger Equation. J. Chem. Phys. 1971; 54, 3534.
15. Bohm D, Vigier JP. Model of the causal interpretation of quantum theory in terms of a fluid with irregular fluctuations. Phys. Rev. 96, 1954; 208-16.
16. Wyatt RE. Quantum wave packet dynamics with trajectories: Application to reactive scattering, J. Chem.Phys, 1999; 111 (10), 4406.
17. Bousquet D, Hughes KH, Micha DA, Burghardt I. Extended hydrodynamic approach to quantum-classical nonequilibrium evolution I. Theory. J. Chem. Phys. 2001; 134.
18. Derrickson SW, Bittner ER. Thermodynamics of Atomic Clusters Using Variational Quantum Hydrodynamics. J. Phys. Chem. 2007; *A*, 111, 10345-10352.
19. Wyatt RE. Quantum Dynamics with Trajectories: Introduction to Quantum Hydrodynamics. Springer, Heidelberg, 2005.
20. Chiarelli P. Can fluctuating quantum states acquire the classical behavior on large scale? arXiv: 1107.4198 [quantum-phys] 2012; submitted for publication on Foundation of Physics.
21. Rumer YB, Ryvkin MS. *Thermodynamics, Statistical Physics, and Kinetics* (Mir Publishers, Moscow, 1980), p. 260.
22. Ibid [21] p. 269.
23. Chiarelli P. The density maximum of He4 at the lambda point modeled by the stochastic quantum hydrodynamic analogy. Int. Arch. 1-3, 1-14 (2012).
24. Anderson JB, Traynor CA, Boghosian BM. J. Chem. Phys. 99 (1), 345 (1993).
25. Chiarelli P. “The Classical Mechanics from the quantum equation”, Phys. Rew. & Res. Int., 3(1) (2013) pp.1-9.

**NOMENCLATURE**

| | |
|---|---|
| n : squared wave function modulus | $l^{-3}$ |
| *S* : action of the system | $m^{-1}\ l^{-2}\ t$ |
| *m* : mass of structureless particles | m |
| $\hbar$ : Plank’s constant | $m\ l^{2}\ t^{-1}$ |
| *c* : light speed | $l\ t^{-1}$ |
| $k_b$ : Boltzmann’s constant | $m\ l^{2}\ t^{-2}$/°K |
| $\Theta$ : Noise amplitude | °K |
| *H* : Hamiltonian of the system | $m\ l^{2}\ t^{-2}$ |
| $V$ : potential energy | $m\, l^{2}\, t^{-2}$ |
| $V_{qu}$ : quantum potential energy | $m\, l^{2}\, t^{-2}$ |
| $\eta$ : Gaussian noise of WFMS | $l^{-3}\, t^{-1}$ |
| $\lambda_{c}$ : correlation length of squared wave function modulus fluctuations | $l$ |
| $\lambda_{L}$ : range of interaction of non-local quantum interaction | $l$ |
| $G(\lambda)$ : dimensionless correlation function (shape) of WFMS fluctuations | pure number |
| $\underline{\mu}$: WFMS mobility form factor | $m^{-1}\, t\, l^{-6}$ |
| $\mu$ = WFMS mobility constant | $m^{-1}\, t$ |

## Appendix A

*Pseudo-Gaussian WFM*

If a system admits the large-scale classical dynamics, the WFM cannot acquire an exact Gaussian shape because it would bring to an infinite quantum non-locality length.

In section 3. we have shown that for $h$ < 3/2 (when the WFM decreases slower than a Gaussian) a finite quantum length is possible.

The Gaussian shape is a physically good description of particle localization but irrelevant deviations from it, at large distance, are decisive to determine the quantum non-locality length.

For instance, let's consider the pseudo-Gaussian function type

$$\mathrm{n} = \mathrm{n}_0 \, exp[-\frac{(q-\underline{q})^2}{\underline{\Delta q}^2 [1+[\frac{(q-\underline{q})^2}{\Lambda^2 f(q-\underline{q})}]]}] \qquad \text{(A.1)}$$

where $f(q-\underline{q})$ is an opportune regular function obeying to the conditions

$$\Lambda^2 f(0) >> \underline{\Delta q}^2 \text{ and } \quad lim_{|q-\underline{q}|\to\infty} f(q-\underline{q}) << \frac{(q-\underline{q})^2}{\Lambda^2} . \qquad \text{(A.2)}$$

For small distance $(q-\underline{q})^2 << \Lambda^2 f(q-\underline{q})$ the above WFM is physically indistinguishable from a Gaussian, while for large distance we obtain the behavior

$$lim_{|q-\underline{q}|\to\infty} \mathrm{n} = \mathrm{n}_0 \, exp[-\frac{\Lambda^2 f(q-\underline{q})}{\underline{\Delta q}^2}] . \qquad \text{(A.3)}$$

For instance, we may consider the following examples

1) $f(q-\underline{q}) = 1$

$$lim_{|q-\underline{q}|\to\infty}\, n = n_0\, exp[-\frac{\Lambda^2}{\underline{\Delta q^2}}]\;; \qquad (A.4)$$

2) $f(q-\underline{q})$ = 1+ |q - $\underline{q}$|

$$lim_{|q-\underline{q}|\to\infty}\, n = n_0\, exp[-\frac{\Lambda^2\,|q-\underline{q}|}{\underline{\Delta q^2}}]\;; \qquad (A.5)$$

3) $f(q-\underline{q}) = 1 + ln[1+|q-\underline{q}|^h] \approx ln[|q-\underline{q}|^g]$ (0< $g$ <2)

$$lim_{|q-\underline{q}|\to\infty}\, n \approx n_0\, |q-\underline{q}|^{-g\frac{\Lambda^2}{\underline{\Delta q^2}}}\;; \qquad (A.6)$$

4) $f(q-\underline{q}) = 1+|q-\underline{q}|^g$ (0< $g$ <2)

$$lim_{|q-\underline{q}|\to\infty}\, n = n_0\, exp[-\frac{\Lambda^2\,|q-\underline{q}|^g}{\underline{\Delta q^2}}] \qquad (A.7)$$

All cases (1-4) lead to a finite quantum non-locality length $\lambda_q$ .

In the case (4) the quantum potential for $|q-\underline{q}|\to\infty$ reads:

$$\begin{aligned} lim_{|q-\underline{q}|\to\infty} V_{qu} &= lim_{|q-\underline{q}|\to\infty} -(\frac{\hbar^2}{2m})|\psi|^{-1}\,\nabla_q \bullet \nabla_q\,|\psi| \\ &= -(\frac{\hbar^2}{2m})[\frac{\Lambda^4 g^2 (q-\underline{q})^{2(h-1)}}{(2\underline{\Delta q^2})^2} - \frac{\Lambda^2 g(g-1)(q-\underline{q})^{g-2}}{\underline{2\Delta q^2}}] \end{aligned} \qquad (0<g<2) \quad (A.8)$$

leading, for 0< $g$ <2, to the quantum force

$$lim_{|q-\underline{q}|\to\infty} -\nabla_q V_{qu} = (\frac{\hbar^2}{2m})[\frac{\Lambda^4 g^2 (2g-1)(q-\underline{q})^{2g-3}}{(2\underline{\Delta q^2})^2} - \frac{\Lambda^2 g(g-1)(g-2)(q-\underline{q})^{g-3}}{\underline{2\Delta q^2}}] \qquad (A.9)$$

that for $g$ < 3/2 gives $lim_{|q-\underline{q}|\to\infty} -\nabla_q V_{qu} = 0$ ,

It is interesting to note that for $g$ =2

$$|\psi| = n^{1/2} = n_0^{1/2}\, exp[-\frac{(q-\underline{q})^2}{\underline{2\Delta q^2}}] \quad \text{(linear case)} \qquad (A.10)$$

the quantum potential is quadratic

$$lim_{|q-\underline{q}|\to\infty} V_{qu} = -(\frac{\hbar^2}{2m})[\frac{(q-\underline{q})^2}{(\underline{\Delta q^2})^2} - \frac{1}{\underline{\Delta q^2}}], \qquad (A.11)$$

and the quantum force is linear and reads

$$lim_{|q-\underline{q}|\to\infty} - \nabla_q V_{qu} = (\frac{\hbar^2}{2m})[\frac{2(q-\underline{q})}{(\underline{\Delta q^2})^2}] \qquad (A.12)$$

The linear form of the force exerted by the quantum potential leads to the ballistic expansion (variance that grows linearly with time) of the free Gaussian quantum states.

## Appendix B

Even if the relation between the SQHA noise fluctuations amplitude $\Theta$ and the temperature T of an ensemble of particles is not T = $\Theta$ *tout court* (see ref. [20]) it can be easily recognized that when we cool a system toward the absolute zero (with steps of equilibrium) also the noise amplitude $\Theta$ reduces to zero since the energy fluctuations of the system must vanish. Hence, we can infer that when the (mechanical or thermodynamic) temperature T is lowered also the WFM noise amplitude $\Theta$ decreases.